\documentclass[reprint, superscriptaddress, secnumarabic, amssymb, nobibnotes, aps, pra]{revtex4-1}

\usepackage{graphicx}
\usepackage{subfiles}
\usepackage{epstopdf}
\usepackage[T1]{fontenc}
\usepackage[latin9]{inputenc}
\usepackage{amsbsy}
\usepackage{gensymb}

\usepackage[T1]{fontenc}
\usepackage[latin9]{inputenc}
\usepackage{amsmath}
\usepackage{amssymb}
\usepackage{bbm}
\usepackage{physics}
\usepackage{braket}
\usepackage{xcolor}
\usepackage{ wasysym }
\allowdisplaybreaks
\usepackage{graphicx}
\usepackage[colorlinks=true]{hyperref}  
\hypersetup{
    bookmarks=true,         
    unicode=false,          
    pdftoolbar=true,        
    pdfmenubar=true,        
    pdffitwindow=false,     
    pdfstartview={FitH},    
    pdftitle={Superconducting properties of ScS},    
    pdfauthor={Arushi et al.},     
    pdfsubject={},   
    pdfcreator={},   
    pdfproducer={}, 
    pdfkeywords={} {} {}, 
    pdfnewwindow=true,      
    colorlinks=true,       
    linkcolor=blue, 
    citecolor=blue,        
    filecolor=magenta,      
    urlcolor=blue           
} 
\usepackage[normalem]{ulem}

\newcommand{\equref}[1]{Eq.~(\ref{#1})}

\newcommand{\figref}[1]{Fig.~\ref{#1}}
\newcommand{\refcite}[1]{Ref.~\onlinecite{#1}}

\newcommand{\tableref}[1]{Table~\ref{#1}}

\newcommand{\diff}{\mathrm{d}}

\renewcommand{\approx}{\simeq}

\renewcommand{\vec}[1]{\boldsymbol{#1}}

\makeatletter
\def\maketitle{
\@author@finish
\title@column\titleblock@produce
\suppressfloats[t]}
\makeatother

\begin{document}
\title{\textrm{Time-reversal-symmetry Breaking in the Superconducting State of ScS}}
\author{Arushi}
\affiliation{Department of Physics, Indian Institute of Science Education and Research Bhopal, Bhopal, 462066, India}
\author{R. K. Kushwaha}
\affiliation{Department of Physics, Indian Institute of Science Education and Research Bhopal, Bhopal, 462066, India}
\author{D.~Singh}
\affiliation{ISIS Facility, STFC Rutherford Appleton Laboratory, Didcot OX11 0QX, United Kingdom}
\author{A.~D.~Hillier}
\affiliation{ISIS Facility, STFC Rutherford Appleton Laboratory, Didcot OX11 0QX, United Kingdom}
\author{M.~S.~Scheurer}
\affiliation{Institute for Theoretical Physics, University of Innsbruck, A-6020 Innsbruck, Austria}
\author{R.~P.~Singh}
\email[]{rpsingh@iiserb.ac.in}
\affiliation{Department of Physics, Indian Institute of Science Education and Research Bhopal, Bhopal, 462066, India}
\begin{abstract}
\begin{flushleft}
\end{flushleft}

We have studied the electronic properties of ScS, a transition-metal monochalcogenide with rocksalt crystal structure, using magnetization, specific heat, transport, and muon spin rotation/relaxation ($\mu$SR) measurements. All measurements confirm the bulk superconducting in ScS with a transition temperature of $T_{C}$ = 5.1(5) K. Specific heat together with transverse-field $\mu$SR measurements indicate a full gap, while our zero-field $\mu$SR study reveals the presence of spontaneous static or quasi-static magnetic fields emerging when entering the superconducting state. We discuss various possible microscopic origins of the observed time-reversal-symmetry breaking. As none of them can be readily reconciled with a conventional pairing mechanism, this introduces ScS as a novel candidate material for unconventional superconductivity.
\end{abstract}

\maketitle
The study of unconventional superconductors \cite{Sigrist}, which go beyond the BCS theory, is a crucial pillar of modern condensed-matter research and involves a broad range of material classes, ranging from heavy fermion systems \cite{HFReview}, high T$_{C}$ superconductors \cite{HT1,HT2}, to iron-based systems \cite{IB1,IB2}, and more recently moir\'e superlattices \cite{TBG}, just to name a few. This research is driven by the potential of these phases for applications and by fundamental scientific questions, such as understanding their pairing mechanism, identifying unifying physical similarities across chemically rather different sets of materials, and finding ways to probe their microscopic physics. 

However, even the identification of unconventional pairing is a challenging endeavor: while the presence of nodes in the gap function, which can be protected by symmetry in an unconventional state, is a good indication for unconventional pairing, there are also more subtle unconventional pairing states with a full and approximately isotropic gap \cite{Sigrist}. In this case, phase sensitive techniques are required, with one example given by studying the disorder sensitivity of the pairing state \cite{Disorder1,Disorder2,Disorder3,Disorder4}. 
Another phase-sensitive identification is the observation of spontaneous internal magnetic fields at the superconducting transition, indicating that the superconducting order parameter breaks time-reversal symmetry \cite{TRS1,TRS2,TRS3,HF1}. Time-reversal-symmetry-breaking superconductivity is particularly interesting since it is rare in nature, the underlying pairing mechanism must involve more than the conventional electron-phonon coupling \cite{SheurerGenRelation}, and due to its potential for technological applications, e.g., for the stabilization of topological edge modes \cite{topo0}, and possibly also for the realization \cite{DiodeThSgt,DiodeTh} of zero-field superconducting diodes \cite{DiodeExp}.

Binary transition metal arsenides (TMA) where TM represents a transition metal and A can be any element from the carbon, pnictogen or chalcogen group, have been widely studied as they provide an exciting family of candidate materials where a range of exotic features has been observed over the years. For instance, NbC, TaC, MoC, VC, and CrC \cite{NbC_TaC,MoC,VC_CrC} exhibit superconductivity together with non-trivial topological band structure. 
These compounds crystallize in a centrosymmetric cubic structure known as rock salt structure. Motivated by the interest in and exciting properties of these materials, we study ScS, which is isostructural to the above-mentioned compounds.

\begin{figure*} 
\centering
\includegraphics[width=2.0\columnwidth, origin=b]{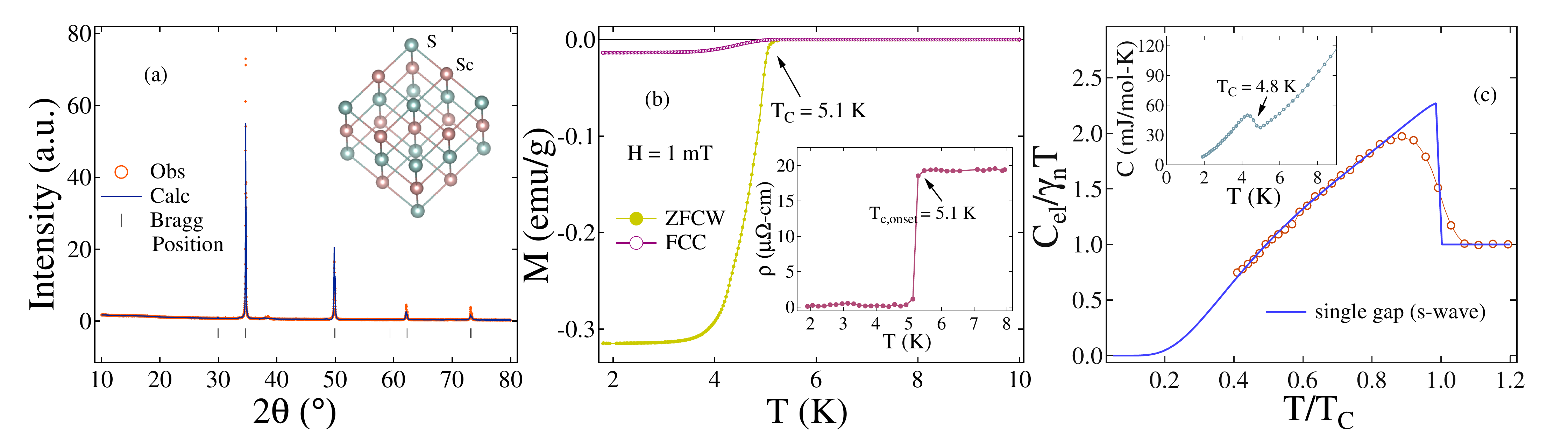}
\caption{\label{Fig1:XRD} a) Powder XRD of ScS obtained at room temperature, shown by open red circles. The solid blue line is the Rietveld refinement whereas the black bars show Bragg reflection peaks. Inset: Crystal structure of ScS. b) Temperature dependence of the magnetic moment shows T$_{C}$ at 5.1 K for ScS. The data were collected in an applied field of 1 mT via ZFCW and FCC protocols. c) Normalized specific heat, $C_{\text{el}}/\gamma_{n}T$ is fitted using single gap $s$-wave model represented by the solid blue line. Inset: Temperature dependence of the total specific heat in zero-field exhibiting T$_{C}$ at 4.8(1) K.}
\end{figure*}

In this work, we report the macroscopic and microscopic study of superconducting properties in ScS by means of magnetization, resistivity, specific heat, and muon spin rotation and relaxation ($\mu$SR) measurements. Superconductivity in ScS has been known for a long time \cite{SCScS1,SCScS2}; however, a detailed study of the superconducting properties has not yet been performed. All of our measurements confirm bulk superconductivity with a transition temperature of 5.1(5) K. Transverse-field (TF) $\mu$SR indicates  a nodeless, approximately isotropic superconducting gap structure together with a slightly enhanced gap to critical temperature ratio compared to weak-coupling BCS theory. 
Zero-field (ZF) $\mu$SR measurements reveal time-reversal symmetry breaking on entering the superconducting state which makes ScS the first member of the rocksalt family to exhibit this exotic feature.

\textit{Thermodynamics and transport---}The polycrystalline sample of ScS was prepared by arc melting both the constituent elements on a water cooled cooper hearth under an argon gas atmosphere. To get structural information, powder X-ray diffraction (XRD) on well grounded powder was performed at 300 K using a PANalytical diffractometer equipped with CuK$_\alpha$ radiation ($\lambda$ = 1.5406 \AA). ScS adopts a cubic (NaCl defect) crystal structure, see inset of \figref{Fig1:XRD}(a), with space group Fm--3m (No.~225),  which was determined by the Rietveld refinement of room temperature powder XRD data shown in the main panel of \figref{Fig1:XRD}(a). The lattice constant is a = 5.172(3) \text{\AA}. To perform magnetization, electrical resistivity and specific heat measurements, SQUID (MPMS 3, Quantum Design) and PPMS were used. 

Magnetization $M$ measurements collected in an applied field of $1\,\textrm{mT}$ via zero-field cooled warming (ZFCW) and field cooled cooling (FCC) modes confirmed the bulk nature of superconductivity in ScS. As can be seen in \figref{Fig1:XRD}(b), it exhibits a diamagnetic signal at a superconducting transition temperature of T$_{C,onset}$ = 5.10(5)~K, where the electrical resistivity data also shows a zero drop in resistivity [inset of \figref{Fig1:XRD}(b)]. The difference between the diamagnetic signal in FCC and ZFCW indicates the type II nature of superconductivity in ScS. We extract a Meissner superconducting volume fraction close to 100~\% from magnetization measurement. Using $M$ $vs$ field $H$ ($M$ $vs$ $T$) curves at different temperatures (fields) and employing the Ginzburg-Landau relations provided in SI \cite{SI} sec. \ref{Magnetization}, we obtained the lower and upper critical field as H$_{C1}$(0) = 21.0(2)~mT and H$_{C2}$(0) = 0.44(1)~T. Two important length scales, the penetration depth $\lambda_{GL}$(0) and coherence length $\xi_{GL}$(0), are found to be 1077(6) $\text{\AA}$ and 274(3) $\text{\AA}$, respectively. The Ginzburg-Landau parameter $\kappa_{GL}$ = $\lambda_{GL}$(0)/$\xi_{GL}$(0) = 4(1) indicates type II superconductivity in ScS. 
Specific heat measurements at zero-field confirmed bulk superconductivity by exhibiting a jump at T$_{C}$ = 4.8(1) K, which is shown in the inset of \figref{Fig1:XRD}(c). From the total specific heat, the electronic specific heat, C$_{\text{el}}$, can be calculated by subtracting the phononic contribution (see SI \cite{SI} sec. \ref{SH}); its temperature dependence is shown in the main panel of \figref{Fig1:XRD}(c). The dimensionless value of electronic specific heat jump at T$_{C}$, $\frac{\Delta C_{\text{el}}}{\gamma_{n}T_{C}}$ = 1.13, is lower than the  weak-coupling BCS result ($1.43$). 
The temperature dependence of the specific heat below $T_C$ follows more closely a nodeless, isotropic superconducting gap model, see fit in \figref{Fig1:XRD}(c) yielding $\Delta$(0)/k$_{B}$T$_{C}$ = 1.65, than a nodal $p$-wave or $d$-wave model \cite{SI}. 
To estimate the underlying coupling strength $\lambda_{\text{M}}$ of superconductivity, we employed the McMillan model \cite{McMillanMain} and find $\lambda_{\text{M}}$ = 0.61(5) \cite{SI}, which indicates moderately coupled pairing in ScS.
Also taking into account the measured residual resistance, we extract a ratio of BCS coherence length and mean free path of $\xi_{0}/l_{e}$ = 13.13, signalling dirty limit superconductivity. Details regarding all the calculated parameters employing magnetization, electrical resistivity, and specific heat measurements, as well as the fitting relations, are presented in the SI \cite{SI}.

\begin{figure*}
\includegraphics[width=2.0\columnwidth]{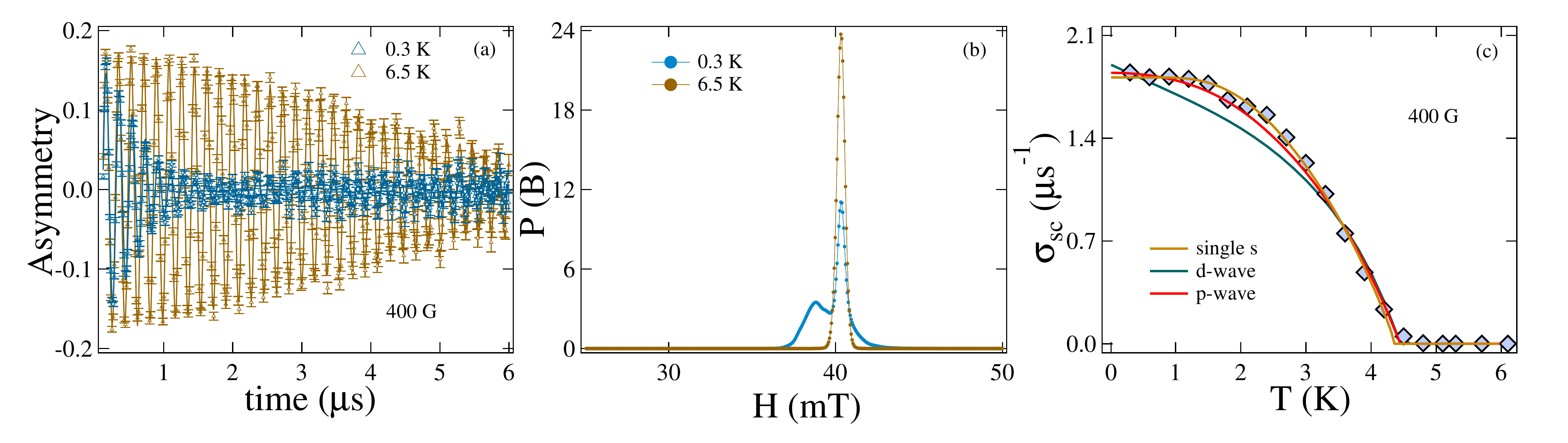}
\caption{\label{Fig2:TF} a) Transverse-field time domain spectra collected at 6.5 K and 0.3 K in an applied magnetic field of 40 mT. The solid lines are fits using \equref{eqn9:TF1}. b) Probability distribution of magnetic field for $T<T_{C}$ and $T>T_{C}$. c) Temperature dependence of $\sigma_{sc}$ (squares), where solid lines represent fitting using \equref{eqn12:BCS1} which employs $s$, $p$, and $d$-wave models.}
\end{figure*}

\textit{$\mu$SR measurements}---Muon spin rotation/relaxation experiments were performed at the ISIS Neutron and Muon facility at the Rutherford Appleton Laboratory, United Kingdom, using a MuSR spectrometer with 64 detectors in both transverse and longitudinal directions. A full description of the muon technique is provided in  \refcite{Muon}. 
TF-$\mu$SR measurements were carried out in FCC mode where the sample was cooled below the transition temperature (to 0.3 K) in the presence of an external magnetic field. The applied magnetic field was well above the lower critical field (H$_{C1}$(0) = 21.0(2) mT) and far below the upper critical critical field which stabilizes the flux line-lattice in the mixed superconducting state. Fig.~\ref{Fig2:TF}(a) shows the asymmetry spectra both above and below the transition temperature. The spectra at 0.3 K exhibit a faster relaxation rate than those at 6.5 K, which is due to the inhomogeneous magnetic field distribution from the flux line lattice. The field distribution in the vortex state at 0.3 K and 6.5 K using maximum entropy algorithm (MaxEnt) is shown in \figref{Fig2:TF}(b). At 0.3 K ($T<T_{C}$), there are two peaks present where one peak corresponds to the applied field sensed by the muons stopping in the sample holder and the other one represents the field distribution due to the flux lattice formation. At T = 6.5~K, only one peak is present at the applied field since ScS is in the normal state. The time domain spectra were best modelled by a sinusoidal oscillating function with a Gaussian relaxation plus a sinusoidal oscillation term for muons hitting the sample holder and a flat background term \cite{TF1,TF2}:
\begin{equation}
\begin{split}
A (t) = A_{1}\exp\left(-\frac{1}{2}\sigma^2t^2\right)\cos(\gamma_\mu B_1t+\phi)\\ + A_{bg}\cos(\gamma_\mu B_{bg}t+\phi) + A_{0},
\label{eqn9:TF1}
\end{split}
\end{equation}
where $A_{1}$, $B_{1}$, and $\sigma$ are the initial asymmetry, internal field, and the Gaussian muon spin relaxation rate belonging to the sample. $\gamma_{\mu}/2\pi$ = 135.5 MHz/T is the muon gyromagnetic ratio and A$_{0}$ is the flat background. $A_{bg}$ and $B_{bg}$ are the asymmetry and field contributions coming from the background when muons hit the sample holder. The relaxation rate corresponding to the superconducting contribution, $\sigma_{\text{sc}}$, can be calculated after subtracting the nuclear magnetic dipolar contribution, $\sigma_{\text{ndip}}$, using the following expression: $\sigma_{\mathrm{sc}} = \sqrt{\sigma^{2} - \sigma_{\mathrm{ndip}}^{2}}$, where $\sigma_{\text{ndip}}$ is assumed to be constant over the entire temperature range; the resulting temperature dependence of $\sigma_{\mathrm{sc}}$ is shown in \figref{Fig2:TF}(c). 
Since $\sigma_{\mathrm{sc}}$ is related to the magnetic penetration depth as $\sigma_{sc} \propto \lambda^{-2}$, it contains information about the form of the superconducting gap structure. Assuming a single, spherical Fermi surface, it holds \cite{TF_sig1,TF_sig2,TF_sig3}
\begin{equation}
\frac{\sigma_{sc}(T)}{\sigma_{sc}(0)} = 1+2\expval{\int_{{|\Delta(T,\hat{\vec{k}})|}}^{\infty}\frac{\partial f}{\partial E}\frac{E\diff E}{\sqrt{E^{2}-\Delta^2(T,\hat{\vec{k}})}}}. \\
\label{eqn12:BCS1}
\end{equation}
Here $\textit{f}$ = [exp($\textit{E}$/$k_{B}T$)+1]$^{-1}$ is the Fermi-Dirac function 
and $\expval{.}$ represents the average over the Fermi surface. $\Delta(T,\hat{\vec{k}})$ = $\Delta_{0}(T) g_{\hat{\vec{k}}}$ is the temperature ($T$) and the directional ($\hat{\vec{k}}$) dependent superconducting gap. Using spherical coordinates with angles $\phi$ and $\theta$, we will consider the cases $g_{\hat{\vec{k}}}=1$, $|\sin(\theta)|$, and $|\cos(2\phi)|$ for a fully gapped $s$-wave, $p$-wave with nodal points, and $d$-wave state with nodal lines, respectively. The temperature dependence of the gap function is approximated by $\Delta_{0}(T) \approx \Delta_{0}(0)\tanh\{1.82(1.018({t^{-1}}-1))^{0.51}\}$ where $\Delta_{0}(0)$ is the magnitude of the superconducting gap at zero temperature.  
As can be seen in Fig.~\ref{Fig2:TF}(c), the data is best captured by the fully gapped $s$-wave model, providing the value of the superconducting gap $\Delta$(0)/k$_{B}$T$_{C}$ = 2.0 which is slightly greater than the gap value estimated from the specific heat (1.43) as well as the standard BCS gap value (1.76), but not inconsistent with the moderately weak coupling constant $\lambda\approx0.6$ extracted above \cite{GapRatio}. 
A small discrepancy between the gap magnitudes obtained from the specific heat and $\mu$SR studies may be due to the lack of specific heat data at low temperatures. To fully understand the gap nature in ScS, further work will be needed such as a detailed study on the Fermi surface of ScS. 

To search for the possible magnetism (static or fluctuating) in ScS, we have performed ZF $\mu$SR measurements as this technique is extremely sensitive to tiny magnetic fields associated with TRS-breaking phases; these measurements were performed in the presence of an active compensation system in order to cancel the stray magnetic field within the range of $0.01\,\textrm{G}$. The time domain spectra were taken above (8.0 K) and below (0.3 K) the transition temperature T$_{C}$ as shown in \figref{Fig3:ZF}(a). There is a significant difference in the relaxation rate observed across T$_{C}$, hinting towards the spontaneous emergence of magnetic fields in the superconducting state. For non-magnetic samples, the depolarization can be best described with the function given below:
\begin{equation}
A(t) = A_{0}G_{\mathrm{KT}}(t)\mathrm{exp}(-\Lambda t)+A_{1} ,
\label{eqn16:ZF2}
\end{equation}
where $\Lambda$ is the Lorentzian relaxation component, A$_{1}$ is the flat background, A$_{0}$ is asymmetry signal coming from the sample; furthermore, G$_{KT}$ is the static Kubo-Toyabe function provided as \cite{ZF_KT}
\begin{equation}
G_{\mathrm{KT}}(t) = \frac{1}{3}+\frac{2}{3}(1-\Delta^{2}t^{2})\mathrm{exp}\left(\frac{-\Delta^{2}t^{2}}{2}\right) ,
\label{eqn15:ZF}
\end{equation}
with $\Delta$ representing the Gaussian muon spin depolarization rate which accounts for the randomly oriented, static nuclear moments experienced at the muon site. The rise in the extracted $\Lambda(T)$, see \figref{Fig3:ZF}(b), in the superconducting state confirms the presence of spontaneous magnetic fields. To exclude the possibility of an impurity induced relaxation, we have performed an additional longitudinal measurement at 0.3 K. A magnetic field of 30 mT was sufficient to decouple the muon spins from the internal magnetic field [\figref{Fig3:ZF}(a)]. It suggests the presence of a static or quasi-static magnetic field.

\begin{figure} 
\centering
\includegraphics[width=1.01\columnwidth, origin=b]{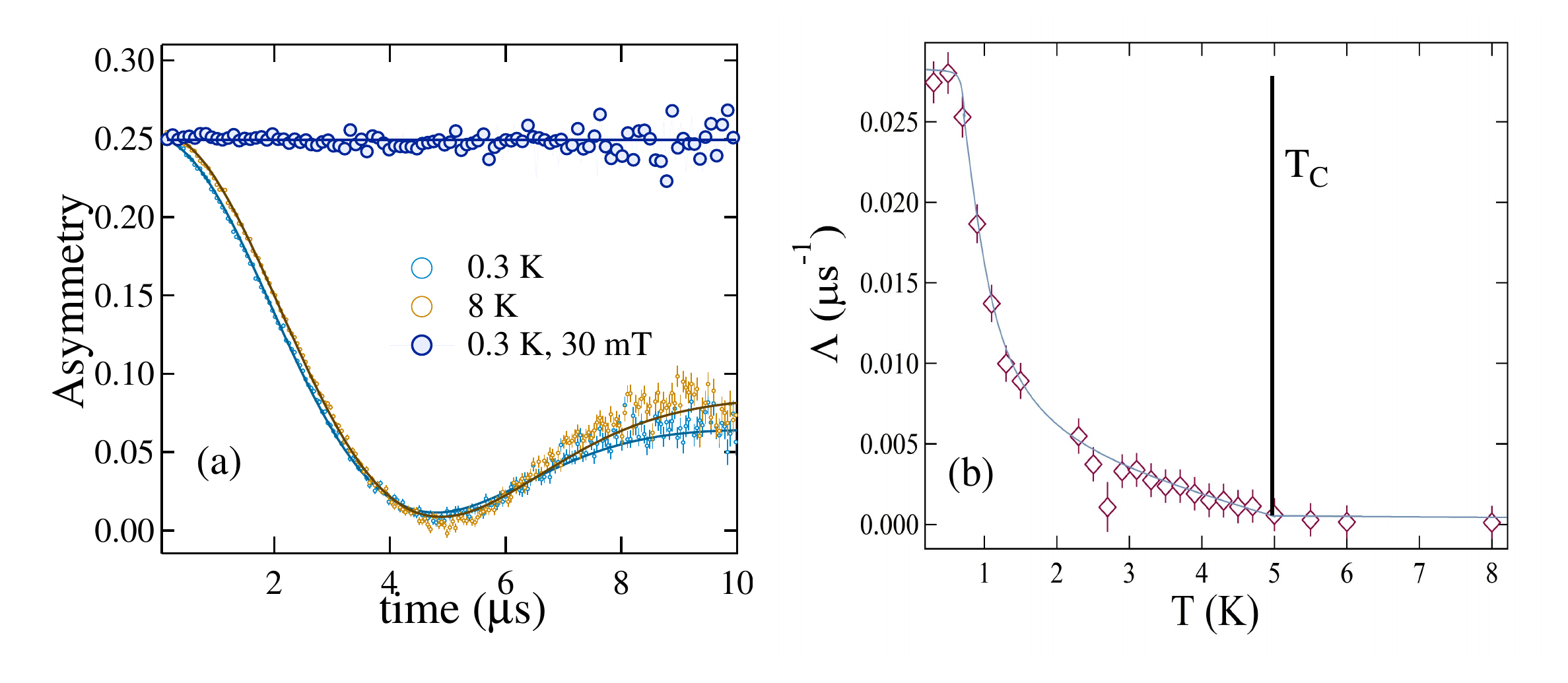}
\caption{\label{Fig3:ZF} a) Asymmetry spectra recorded at two different temperatures, 0.3 K and 8.0 K, in zero magnetic field (small open circles); data at a small longitudinal field (30 mT) are shown by large filled circles.  The solid lines are the corresponding fits to the data using \equref{eqn16:ZF2}. b) Shows the temperature dependence of the relaxation rate, $\Lambda$, featuring an increase below the superconducting transition temperature $T_C$.}
\end{figure}

\textit{Discussion}---Taken together, our measurements of specific heat and penetration depth point towards a fully developed nodeless superconducting gap, while our zero-field $\mu$SR data indicate the emergence of weak magnetic moments below a temperature very close to the superconducting $T_C$.
To explore the implications for the superconducting state in ScS, we will next discuss and critically evaluating four possible microscopic origins, labeled as scenario (i-iv) below, of this phenomenology. 

In \textit{scenario (i)}, we assume that the superconducting phase is reached by a single phase transition, which is natural as there are no indications of multiple superconducting transitions. 
The order parameter must then transform under one of the (10) irreducible representations (IRs) of the normal-state point group $O_h$. 
We further assume that the observed magnetic moments result from the time-reversal-symmetry breaking of the superconducting state itself. As is well-known, this is only possible if it transforms under a two- or higher-dimensional IR---in our case, one of the six IRs $E_g$, $T_{1g}$, $T_{2g}$, $E_u$, $T_{1u}$, $T_{2u}$, leading to a total of $10$ possible time-reversal-symmetry-breaking candidate states \cite{Sigrist}. All of them are required by symmetry to have at least one nodal direction (in some cases even nodal planes) around the $\Gamma$ point. However, first-principle calculations \cite{FirstPrinciple} predict two Fermi surfaces enclosing the $\Gamma$ point completely, such that any of these superconductors is expected to have nodes. While this seems to be at odds with our specific heat and penetration depth data, it is possible that these nodes are not resolved in our measurements. 
As such, we believe that specific heat measurements at lower temperature or more accurate penetration depth studies, e.g., using the tunnel diode resonator technique \cite{TunnelDiodeRes}, might help decide whether this scenario is realized in ScS or not.

Another possibility, \textit{scenario (ii)}, to reconcile broken time-reversal symmetry and a full gap is based on having two consecutive superconducting transitions that are so close that they cannot be resolved. When the dominant repulsive Cooper-channel interactions are \textit{between} the three symmetry-unrelated Fermi sheets of the system, a natural compromise in this `frustrated' situation might be to have non-trivial complex phases of the order parameter between these sheets and a fully gapped $s+i\,s$ state might be realized (see, e.g., the toy model discussion in \cite{LaNi} or the mechanism for time-reversal-symmetry breaking superconductivity in the iron-based superconductors \cite{Andrey,BaKFeAs}); such a state will give rise to magnetic moments \cite{spisFields} that can be detected by the muons. Note, however, that by virtue of transforming under the trivial (and thus one-dimensional) IR of $O_h$, such a state can only be reached by two, possibly very close, phase transitions. We hope that future high-quality crystals will exhibit a sharper phase transition signature in the specific heat, possibly allowing to confirm or rule out this scenario.

One additional complication for both of these scenarios is that we have estimated the coherence length $\xi_0$ to be larger than the mean-free path $l_e$, indicating a significant amount of disorder in the superconductor.  
Therefore, common wisdom \cite{BW,GolubovMazin} would imply that any of the aforementioned unconventional superconducting states should be completely suppressed by impurity scattering. However, more recent theory \cite{DisorderSOCFu,OurDisorderSOC,BrydonScattering,PdTeScattering,Jonathan} has reveal that spin-orbit coupling can protect unconventional pairing states, which is also supported by experiment \cite{Ando2012,Ando2014,Welp}. Moreover, disorder can even be the driving force inducing a time-reversal-symmetry-breaking superconductor, which leads us to \textit{scenario (iii)}: as demonstrated in recent theoretical works on $d$-wave superconductors \cite{Brian,DisorderCuprates}, when two superconducting pairing channels are in close competition, strong disorder can locally induce complex admixtures of these orders leading to local currents, even if the clean sample was in a time-reversal-symmetric state. 
Finally, as pointed out in \cite{SrPtAs}, in a granular sample like the one studied here, a single sign-changing superconducting order parameter can give rise to flux trapping in voids formed by three or more crystallites, defining \textit{scenario~(iv)}.

Importantly, all of these scenarios, (i-iv), require effectively repulsive Cooper-channel interactions, at least for parts of the Fermi surface, and thus necessitate \cite{BrydonRelation,SheurerGenRelation} an unconventional pairing mechanism in the sense that the pairing state cannot be understood in terms of electron-phonon coupling alone. 
The only conceivable picture to explain our observations with a conventional pairing mechanism requires assuming that there are local magnetic moments that are strongly screened in the normal state. At the onset of superconductivity, this screening might be reduced and could lead to the enhancement of the relaxation rate in \figref{Fig3:ZF}(b) below $T_C$; this was recently proposed \cite{TaS2} for $4Hb$-TaS$_2$ where signs of Kondo screening have been observed \cite{KondoScreening}. However, for ScS, we did not find any signs of screening of magnetic moments in the metallic state, rendering this scenario improbable.

\textit{Conclusion and outlook}---We have presented transport, magnetization, specific heat, and $\mu$SR experiments in the superconducting and normal state of ScS, which crystallizes in rocksalt (NaCl) structure. All measurements confirmed the bulk nature of superconductivity, with a transition temperature T$_{C}$ = 4.6 K. We extracted various superconducting and normal-state parameters of ScS. 
Specific heat data and the temperature-dependence of the penetration depth following from TF $\mu$SR measurements are most naturally explained by a fully established superconducting gap. Surprisingly, our zero-field $\mu$SR data reveal time-reversal-symmetry-breaking moments at the onset of superconductivity. We have discussed several possible microscopic origins of these moments, which suggest that the underlying pairing glue cannot arise solely from the electron-phonon coupling, but rather requires repulsive components. As such, our results establish ScS as a system exhibiting complex superconducting properties that deserve further investigation.
In particular, $\mu$SR measurements in high-quality single crystals, specific heat measurements at low temperatures, as well as complementary penetration depth measurements \cite{TunnelDiodeRes} and controlled disorder studies should be able to elucidate the microscopics of superconductivity in ScS.

R.~P.~S.\ acknowledge Science and Engineering Research Board, Government of India for the Core Research Grant CRG/2019/001028. Department of Science and Technology, India (Grant No. SR/NM/Z-07/2015) for the financial support and Jawaharlal Nehru Centre for Advanced Scientific Research (JNCASR) for managing the project. Arushi acknowledges the funding agency, University Grant Commission (UGC) of Government of India for providing SRF fellowship. We thank ISIS, STFC, UK for the beamtime to conduct the $\mu$SR experiments [RB2068032].

\clearpage

\title{Supplementary Information to "Time-reversal-symmetry Breaking in the Superconducting State of ScS"}\label{ExtractParameter}
\maketitle

\subsection{Synthesis and Structural Characterization}\label{XRD}
The starting materials for the preparation of ScS were scandium ingot, and sulphur powder which were weighed in a ratio of 1:1.2 and melted together on a water-cooled copper hearth under an argon gas atmosphere. The as-cast ingots were flipped several times and remelted in order to achieve homogeneity. Rietveld refinement of powder XRD confirmed the crystal structure as cubic and single phase nature within the detection limit of the technique. Other parameters obtained from the refinement, such as atomic positions, cell volume, are summarized in Table \ref{XRD}.
\renewcommand{\thetable}{S\arabic{table}}
\begin{table}[h!]
\caption{Structure parameters of ScS obtained from the Rietveld refinement of XRD}
\label{XRD}
\begin{tabular}{l r} \hline\hline
Structure& Cubic\\
Space group&        F\textit{m}-3\textit{m}\\ [1ex]
Lattice parameters\\ [0.5ex]
a (\text{\AA})&  5.172(3)\\
V$_{\text{Cell}}$ (\text{\AA}$^{3}$)& 138.34(3)
\end{tabular}
\\[1ex]

\begingroup
\setlength{\tabcolsep}{8pt}
\begin{tabular}[b]{c c c c c c}
Atom&  Wyckoff position& x& y& z\\[1ex]
\hline
Sc& 4b& 0.5& 0.5& 0.5\\             
S& 4a& 0& 0& 0\\
[1ex]
\hline
\end{tabular}
\par\medskip\footnotesize
\endgroup
\end{table}

\subsection{Electrical Resistivity}\label{resistivity}

The temperature dependence of electrical resistivity for 1.9 K $\leq$ T $\leq$ 300 K in zero applied field is shown in \figref{S1:Res}. The residual resistivity ratio (RRR = $\rho$(300K)/$\rho$(10K)) is found to be 3.2 which indicates the poor metallic nature and presence of disorder in the system. The normal-state resistivity was analyzed in the framework of the Bloch-Gr\"uneisen (BG) model; more specifically, we write
\setcounter{equation}{0} 
\renewcommand{\theequation}{S\arabic{equation}}
\begin{subequations}\begin{equation}
 \rho(T) = \rho_{0} + \rho_{\text{BG}}(T)
\label{para1}
\end{equation}
where $\rho_{0}$ is the temperature-independent residual resistivity, resulting from scattering due to the presence of defects, and $\rho_{\text{BG}}$ is the BG resistivity, given by \cite{BG1}
\begin{equation}
 \rho_{\text{BG}}(T) = C\left(\frac{T}{\Theta_{D}}\right)^{n}\int_{0}^{\Theta_{D}/T}\frac{x^{n}}{(e^{x}-1)(1-e^{-x})}\mathrm{d}x, 
\label{para2}
\end{equation}\label{BGModel}\end{subequations}
which results from the scattering of electrons and phonons. In \equref{para2}, $\Theta_{D}$ is the Debye temperature, $C$ is a material-dependent prefactor and $n$ takes values from 2 to 5 depending upon the nature of electron scattering \cite{BG2}. The orange solid line in \figref{S1:Res} represents the fit to the data and yields $\rho_0$ = 19.3(1) $\mu\Omega$-cm, $\Theta_{D}$ = 278(5) K, $C$ = 80.7(1) $\mu\Omega$-cm, and $n$ = 3. In order to extract the carrier density for ScS, we have measured the Hall resistivity ($\rho_{xy}$) as a function of magnetic field, which is shown in the inset of \figref{S1:Res}. $\rho_{xy}$ data was fit with a linear line that is shown by blue curve and yields the Hall coefficient $R_{H} = 1.35(9) \times 10^{-10} \ohm$ $m$~T$^{-1}$. The positive value of $R_{H}$ indicates that holes are the dominant carriers in ScS. Using $n$ = $\frac{1}{R_{H} e}$, we find a carrier density of $n$ = 4.6(3)$\times 10^{28} m^{-3}$.

\setcounter{figure}{0}     
\renewcommand{\thefigure}{S\arabic{figure}} 
\begin{figure}[h!]
\includegraphics[width=0.80\columnwidth, origin=b]{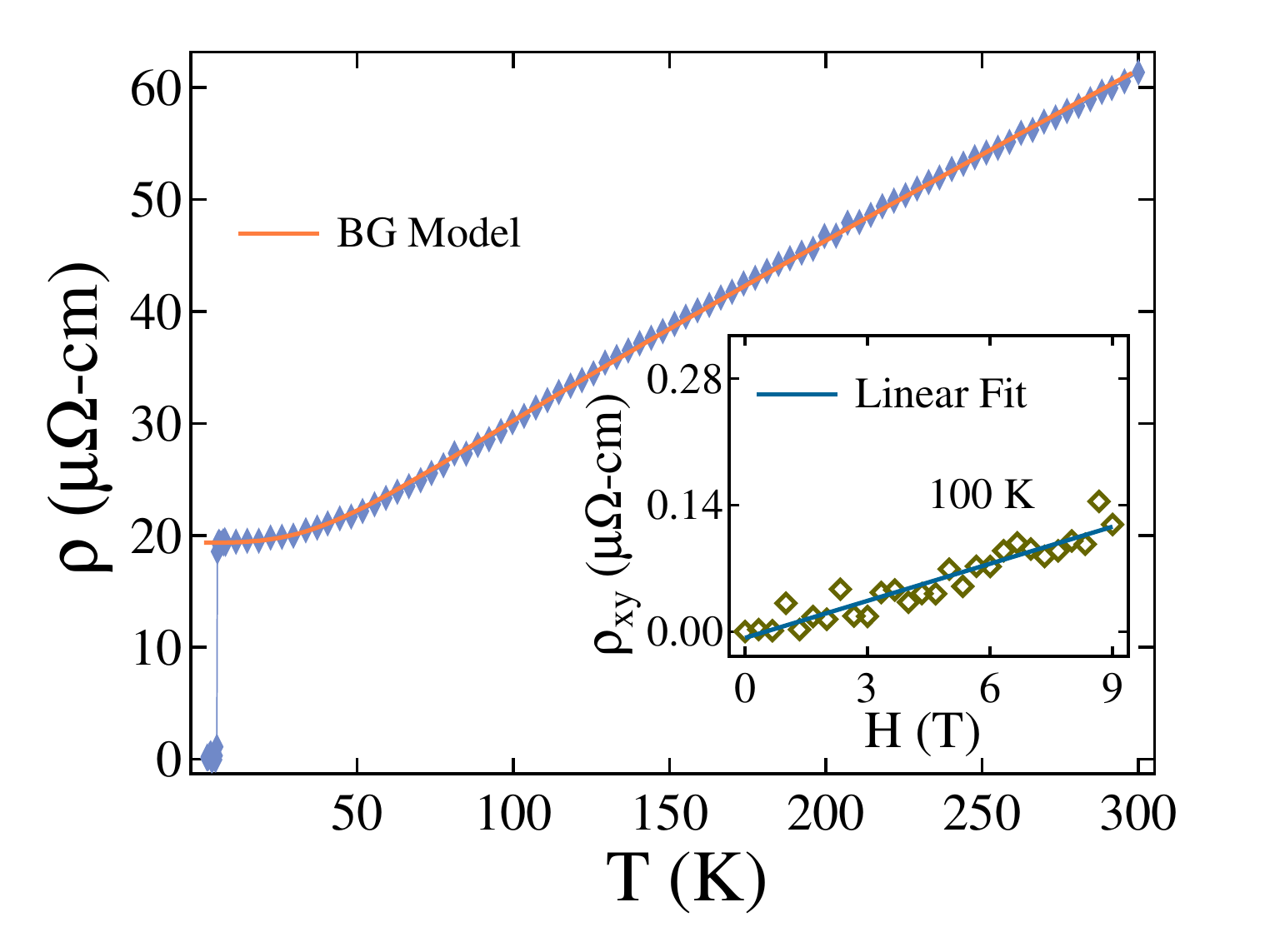}
\caption{\label{S1:Res} Temperature dependence of resistivity in the applied zero field over the temperature range 1.9 K $\leq$ T $\leq$ 300 K. The solid line shows the fit to the BG model in \equref{BGModel}. The lower right inset shows the enlarged view of the $\rho$(T) data, with superconducting transition at T$_{C,onset}$ = 5.1(2) K, and the inset in the top left corner displays the magnetic field dependence of the Hall resistivity.}
\end{figure}

\subsection{Critical Fields}\label{Magnetization}

\begin{figure}
\includegraphics[width=0.80\columnwidth]{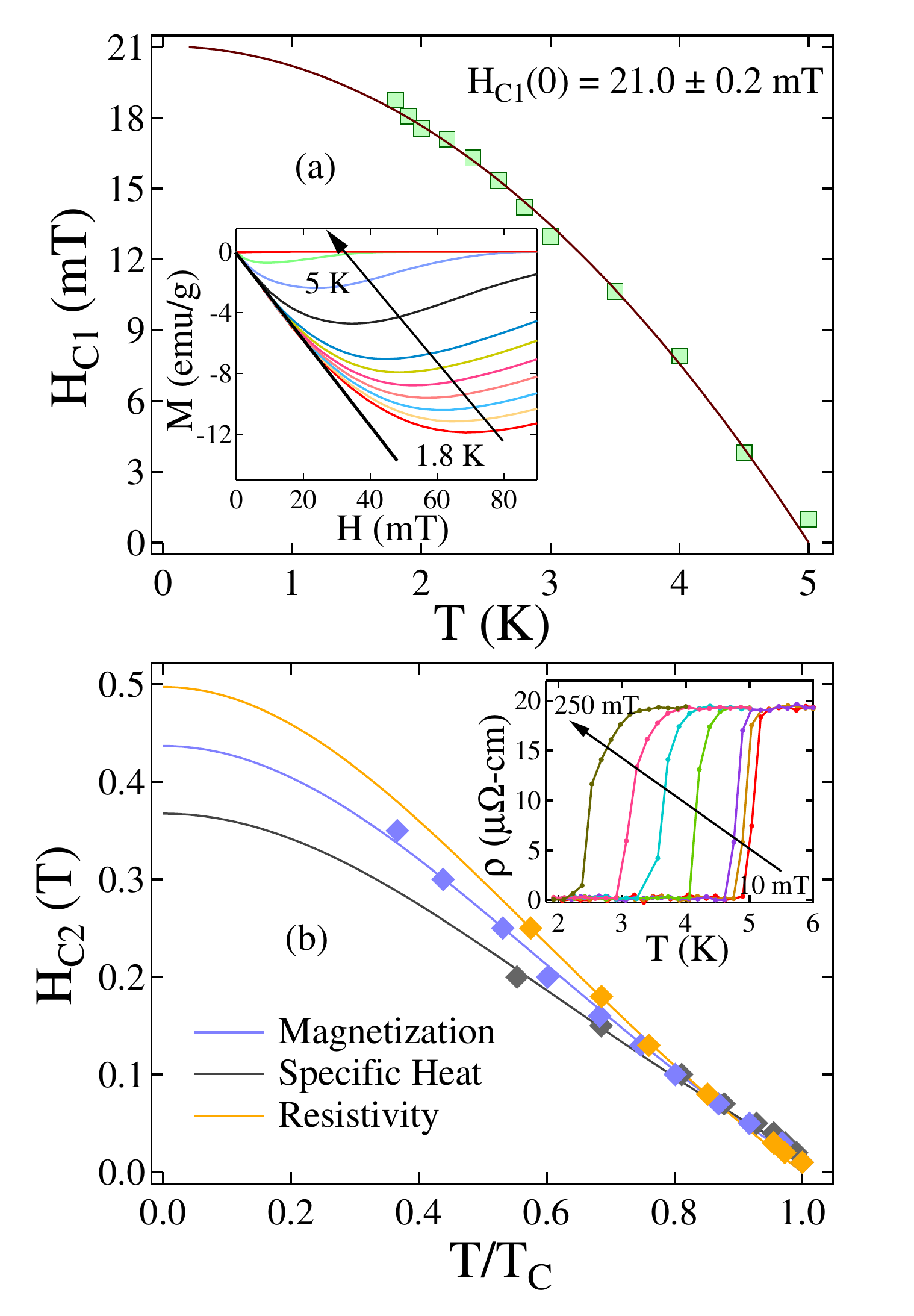}
\caption{\label{S2:Mag} a) Estimated lower critical field, H$_{C1}$, as a function of temperature, where the solid line is a fit according to \equref{eqn1:Hc1}. The inset shows low-field magnetization curves at various temperatures (1.8 $\leq$ T $\leq$ 5 K). b) Temperature dependence of upper critical field obtained from specific heat, AC transport, and magnetization data. The solid lines represent the fits using \equref{eqn2:Hc2}. The inset shows the temperature dependence of the resistivity at different applied fields.}
\end{figure}

To determine the lower critical field, H$_{C1}$, and its temperature dependence, we use the magnetization versus field curves at different temperatures, see inset of \figref{S2:Mag}(a), and define H$_{C1}$ via the field strength at which $M$ deviates from linearity ($M\propto H$). These data points were fitted with the Ginzburg-Landau equation
\begin{equation}
H_{C1}(T)=H_{C1}(0)\left[1-\left(\frac{T}{T_{C}}\right)^{2}\right],
\label{eqn1:Hc1}
\end{equation} 
which yields H$_{C1}$(0) = 21.0(2) mT. 

The upper critical field, $H_{C2}(T)$, was estimated from the field dependence of three different physical quantities: i) the specific heat $C(T,H)$, ii) the magnetization $M(T,H)$, and iii) $\rho(T,H)$; to this end, the transition temperature for magnetization is defined as the temperature that corresponds to the onset of superconductivity and the midpoint of transition temperature in the case of specific heat and resistivity measurements. 
As shown in \figref{S2:Mag}(b), the resulting $H_{C2}(T)$ are roughly the same in the three approaches and, most importantly, all show approximately linear behavior close to $T_C$ and can be fit well with
\begin{equation}
H_{C2}(T) = H_{C2}(0)\left[\frac{(1-t^{2})}{(1+t^2)}\right], \quad t=\frac{T}{T_C}. 
\label{eqn2:Hc2}
\end{equation} 
This provides H$_{C2}$(0) = 0.37(1) T, 0.44(1) T, and 0.49(1) T for specific heat, magnetization, and resistivity, respectively. 

There are two different mechanisms by which a magnetic field interacts with the conduction electrons in a nonmagnetic superconductor and destroys superconductivity. One is orbital pair breaking where the magnetic field couples to the orbital motion of the electron, and the other is the Pauli paramagnetic limiting effect, in which the interaction of the magnetic field is with the electronic spin. 
The orbital limit for an upper critical field is given by the Werthamer-Helfand-Hohenberg (WHH) expression \cite{WHH1,WHH2} 

\begin{equation}
H_{C2}^{orbital}(0) = -\alpha T_{C}\left.\frac{dH_{C2}(T)}{dT}\right|_{T=T_{C}}
\label{eqn3:HHH}
\end{equation}
\\
The initial slope $\frac{-dH_{C2}(T)}{dT} $ at $T = T_{C}$ is estimated to be 0.07(1) T/K. Considering $\alpha$ = 0.693 for dirty limit superconductors, gives the orbital limiting upper critical field $ H_{C2}^{orbital}(0)$ = 0.27 T. The value obtained is less than the value of the critical field at $T/T_{C}$ = 0.55 which suggests that single gap model cannot accurately model this system and multigap effects should be considered. 

Within BCS theory, the value of the Pauli paramagnetic limiting field is given by $ H_{C2}^{P}(0)$ = $c_P \,T_{C} $ where $c_P$ = 1.86 T/K \cite{Pauli1,Pauli2}. It yields $ H_{C2}^{P}(0)$ = 9.8 T, which is much higher than the obtained value of H$_{C2}$(0). The Maki parameter \cite{Maki}, defined as $\alpha_{M} = \sqrt{2}H_{C2}^{orb}(0)/H_{C2}^{P}(0)$, is a measure of the relative strength of Pauli and orbital limits for the upper critical field; we get $\alpha_{M}$ = 0.04, which implies that orbital effects crucially determine H$_{C2}$(0) while the Pauli limiting field is less relevant for ScS.

To determine the Ginzburg-Landau coherence length $\xi_{\text{GL}}$(0), the vale of H$_{C2}$(0) as extracted above from our measurements has been used in the relation $H_{C2}(0) = \frac{\Phi_{0}}{2\pi\xi_{\text{GL}}^{2}}$ \cite{Coh_Leng}, where $ \Phi_{0}$ = 2.07 $\times$ 10$^{-15}$ Tm$^{2}$ is the magnetic flux quantum. It yields $\xi_{\text{GL}}$(0) = 274(3) $\text{\AA}$. In turn, the respective values of $\xi_{\text{GL}}$(0) = 274(3) $ \text{\AA}$ and H$_{C1}$(0) = 21.0(2) mT are used to calculate the Ginzburg-Landau penetration depth, $\lambda_{\text{GL}}$, using  \cite{pene}
\begin{equation}
H_{C1}(0) = \frac{\Phi_{0}}{4\pi\lambda_{GL}^2(0)}\left(\mathrm{ln}\frac{\lambda_{GL}(0)}{\xi_{GL}(0)}+0.12\right).   
\label{eqn4:PD}
\end{equation} 
This yields $\lambda_{\text{GL}}$(0) = 1077(6) $\text{\AA}$. The Ginzburg-Landau parameter $\kappa_{\text{GL}} = \frac{\lambda_{\text{GL}}(0)}{\xi_{\text{GL}}(0)}$, which distinguishes between type I and type II superconductivity, is evaluated to be 4.0(1). Being significantly larger than $1/\sqrt{2}$, this value demonstrates the type II nature of the superconductivity in ScS. The thermodynamic critical field H$_{C}$ has been estimated by using the relation \cite{pene} $H_{C1}(0)H_{C2}(0) = H_{C}^2\mathrm{ln}\kappa_{\text{GL}}$, which yields H$ _{C} $ = 82(2) mT.

\subsection{Specific Heat}\label{SH}

\begin{figure}
\includegraphics[width=0.80\columnwidth]{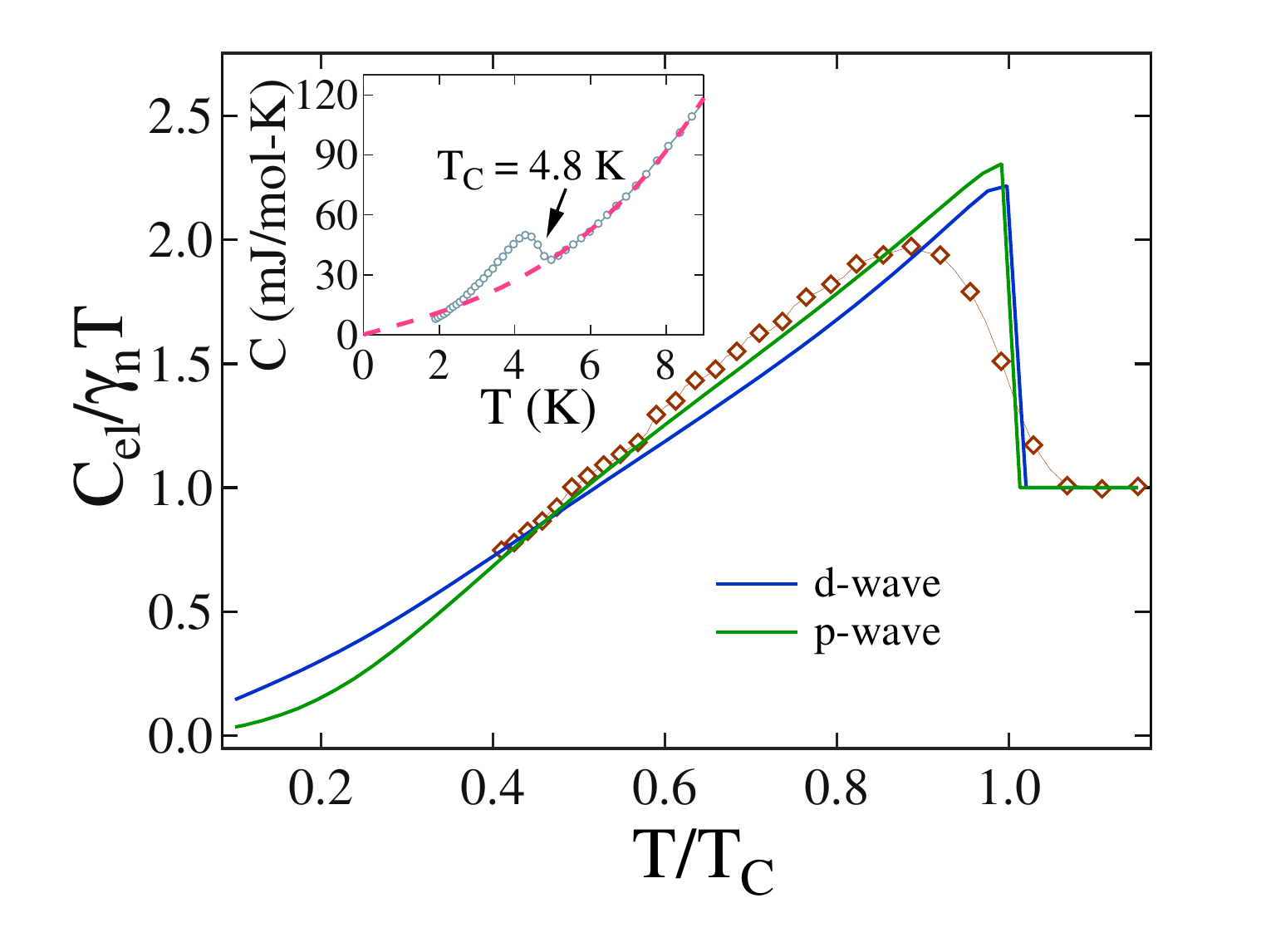}
\caption{\label{S3:SH} Solid lines show fits to normalized specific heat, $C_{\text{el}}/\gamma_{n}T$ data using $p$ wave and $d$ wave models.}
\end{figure}

The temperature dependence of the specific heat $C(T)$ in the temperature range 1.9 K to 9.0 K in the zero applied field is shown in the inset of \figref{S3:SH}. Above T$_{C}$, the specific heat can be well described with the expression
\begin{equation}  
\frac{C}{T}=\gamma_{n}+\beta_{3} T^{2},
\label{eqn5:SH1}    
\end{equation}
where $\gamma_{n}$ is the Sommerfeld coefficient, which represents the electronic contribution, and $\beta_{3}$ describes the phononic contribution. The fit to the data yields: $ \gamma_{n}$ = 5.17(5) mJmol$^{-1}$K$^{-2}$, and $\beta_{3}$ = 98.6(9) $\mu$Jmol$^{-1}$K$^{-4}$. The Sommerfeld coefficient, $\gamma_{n}$, is used to estimate the density of states at the Fermi level, $D_{C}(E_{\mathrm{F}})$, via the free-electron relation, $\gamma_{n}= \left(\frac{\pi^{2}k_{B}^{2}}{3}\right)D_{C}(E_{\mathrm{F}})$, where k$_{B}$ = 1.38$\times$10$^{-23}$ J K$^{-1}$. This gives $D_{C}(E_{\mathrm{F}})$ = 2.19(2) states eV$^{-1}$f.u$^{-1}$. Debye temperature, $\Theta_{D}$ is estimated via $\Theta_{D}= \left(\frac{12\pi^{4}RN}{5\beta_{3}}\right)^{\frac{1}{3}}$, where R = 8.314 J mol$ ^{-1} $K$ ^{-1} $ is a gas constant, $N$ is the number of atoms per formula unit, and $\beta_{3}$ = 98.6(9) $\mu$Jmol$^{-1}$K$^{-4}$ is the Debye constant extracted from \equref{eqn5:SH1}. After employing the values, we obtained $\Theta_{D}$ = 340(3) K. The electron-phonon coupling parameter, $\lambda_{\text{M}}$ can be estimated using McMillan's model \cite{McMillan} which relates $\Theta_{D}$ and T$_{C}$ as given below:
\\
\begin{equation}
\lambda_{\text{M}} = \frac{1.04+\mu^{*}\mathrm{ln}(\Theta_{D}/1.45T_{C})}{(1-0.62\mu^{*})\mathrm{ln}(\Theta_{D}/1.45T_{C})-1.04 }
\label{eqn6:Lambda}
\end{equation}
\\
where $\mu^{*}$ is the screened Coulomb repulsion and the typical value of 0.13 will be used here. From the values of $\Theta_{D}$ = 340(3) K and T$_{C}$ = 4.8(1) K, we obtain $ \lambda_{\text{M}} $ = 0.61(5), which classifies ScS as a moderately coupled superconductor. 

Having extracted the values of the Sommerfeld coefficient $\gamma_n$, carrier density $n$ and residual resistance $\rho_{0}$, we can estimate the ratio of mean free path and coherence length $\xi_{0}$/$l_{e}$. Based on the Drude model, the mean free path is given by $l_{e} = v_{F}\tau$ where $\tau$ is the scattering time and is defined as $\tau^{-1}$ = $ne^{2}\rho_{0}/m^{*}$. The effective mass $m^{*}$ is given by the expression: $m^{*}$ = ($\hbar k_{F})^{2}\gamma_{n}$/$\pi^{2}nk_{B}^{2}$ where $k_{F}$ = (3$\pi^2n$)$^{1/3}$ when considering a spherical Fermi surface. We find $k_{F}$ = 1.11(2) $\text{\AA}^{-1}$ and $m^{*}$ = 5.2(5) $m_{e}$ employing the values of $n$ and $\gamma_{n}$ 
obtained above. The Fermi velocity $v_{F} = \hbar k_{F}/m^{*}$ is found to be 2.5(2) $\times$ 10$^{5}$ m/s and the mean free path $l_{e}$ is 51(4) $\text{\AA}$. Within the BCS framework, the coherence length is given via the expression: $\xi_{0}$ = 0.18$\hbar v_{F}$/$k_{B}T_{C}$ and by substituting the values of $v_{F}$ and $T_{C}$, we obtained $\xi_{0}$ = 667(59) $\text{\AA}$. All parameters are listed in \tableref{AllParameters} together with all other parameters extracted from our measurements. The ratio $\xi_{0}$/$l_{e}$ indicates the presence of significant disorder in the superconducting phase (consistent with the rather low RRR value discussed above).

\begin{table}[tb]
\caption{Superconducting and normal-state parameters of ScS}
\begingroup
\setlength{\tabcolsep}{12pt}
\label{AllParameters}
\begin{tabular}{c c c c} 
\hline\hline
Parameters & Units & ScS\\ [1ex]
\hline
$T_{C}$& K& 5.1\\             
$H_{C1}(0)$& mT& 21.03\\                       
$H_{C2}(0)$& T& 0.44\\
$H_{C2}^{P}(0)$& T& 9.8\\
$H_{C2}^{Orb}(0)$& T& 0.26\\
$\xi_{GL}$& \text{\AA}& 274\\
$\lambda_{GL}$& \text{\AA}& 1077\\
$k_{GL}$& &4\\
$\gamma_{n}$&  mJ mol$^{-1}$ K$^{-2}$& 5.16\\   $\Theta_{D}$& K& 340\\
$\Delta C_{el}/\gamma_{n}T_{C}$&   &1.13\\
$\xi_{0}/l_{e}$&   &13.13\\
$\xi_{0}$& \text{\AA}& 667\\
$l_{e}$& \text{\AA}& 51\\
$v_{F}$& 10$^{5}$m s$^{-1}$& 2.46\\
$n$& 10$^{28}$m$^{-3}$& 4.59\\
$m^{*}$/m$_{e}$&  &5.21 \\
[1ex]
\hline\hline
\end{tabular}
\endgroup
\end{table}

From total specific heat, electronic specific heat, C$_{\text{el}}$, can be calculated subtracting the phononic contribution, C$_{\text{ph}}  = \beta_{3} T^{3}$; its temperature dependence is shown in the main panel of \figref{S3:SH}. The value of the electronic specific heat jump at the superconducting transition is found to be $\frac{\Delta C_{\text{el}}}{\gamma_{n}T_{C}}$ = 1.13, which is smaller than the BCS weak coupling result ($1.43$). A reduced magnitude of the jump has been observed in specific heat studies of several superconductors. These studies point toward the contribution of impurities, inhomogeneity, or the existence of regions which do not take part in superconductivity \cite{lowjump}. In our sample, XRD confirms the presence of a single phase, which disfavors this explanation. Another possible origin of the reduced jump is a more complex momentum and Fermi-surface dependent, gap structure distinct from the BCS model.

The electronic specific heat in the superconducting state is calculated from the normalized entropy expression, which can be written as \cite{SH1,SH2,SH3}
    \begin{equation}
    \frac{S(t)}{\gamma_{n}T_{C}} = -\frac{6}{\pi^2}\int_{0}^{\infty}\hspace{-0.2em}\diff\epsilon\,\,[f \ln f + (1-f) \ln (1-f)],  \label{NormalizedEntropyExpr}
    \end{equation}
which depends only on the reduced temperature $t = T/T_{C}$. In \equref{NormalizedEntropyExpr}, $f = [\exp(\textit{E}/k_{B}T)+1]^{-1}$ is the Fermi function and E($\epsilon$) = $\sqrt{\epsilon^{2}+\Delta_{k}^{2}}$ where $\epsilon$ is the relative energy(measured in units of $T_C$). $\Delta_{k}$ represents the temperature and angle dependence similar to as described in the main text. The normalized entropy S is related to the normalized electronic specific heat by 
\begin{equation}
\frac{C_{\text{el}}}{\gamma_{n}T_{C}} = t\frac{d(S/\gamma_{n}T_{C})}{dt} \\
\label{eqn8:BCS2}
\end{equation}

We have analyzed the temperature dependence of the normalized specific heat with the aforementioned models where the fitting is represented by the solid lines shown in \figref{S3:SH}. As can be seen, the nodal $p$- and $d$-wave ans\"atze can not provide a fit in the measured temperature range. While specific heat can in principle be used to distinguish between single- and multi-gap $s$-wave behavior, the available temperature range does not allow to distinguish reliably between different $s$-wave pairing scenarios.


\begin{thebibliography}{References}
\bibitem{Sigrist} M. Sigrist and K. Ueda, Rev. Mod. Phys. \textbf{63}, 239 (1991).

\bibitem{HFReview} B.D. White, J.D. Thompson, M.B. Maple, Physica C: Superconductivity and its Applications, 514, 246-278 (2015).

\bibitem{HT2} D. J. Van Harlingen, Rev. Mod. Phys. 67, 515 (1995).

\bibitem{HT1} S. Sachdev, Rev. Mod. Phys. 75, 913 (2003).

\bibitem{IB1} F. Wang and D.-H. Lee, Science 332, 200-204 (2011).

\bibitem{IB2} X. Chen, P. Dai, D. Feng, T. Xiang, and F.-C. Zhang, Nat. Sci. Rev.  1, 3 (2014).

\bibitem{TBG} Y. Cao, V. Fatemi, S. Fang, K. Watanabe, T. Taniguchi, E. Kaxiras, P. and Jarillo-Herrero, Nature \textbf{556}, 43-50 (2018).


\bibitem{Disorder1} A. P. Mackenzie, R. K. W. Haselwimmer, A. W. Tyler, G. G. Lonzarich, Y. Mori, S. Nishizaki, and Y. Maeno, Phys. Rev. Lett. \textbf{80}, 161 (1998).

\bibitem{Disorder2} F. Rullier-Albenque, H. Alloul, and R. Tourbot, Phys. Rev. Lett. \textbf{91}, 047001 (2003).

\bibitem{Disorder3} J. Li, Y. Guo, S. Zhang, S. Yu, Y. Tsujimoto, H. Kontani, K. Yamaura, and E. Takayama-Muromachi, Phys. Rev. B \textbf{84}, 020513(R) (2011).

\bibitem{Disorder4} E. H. Krenkel, M. A. Tanatar, M. Konczykowski, R. Grasset, E. I. Timmons, S. Ghimire, K. R. Joshi, Y. Lee, Liqin Ke, S. Chen, C. Petrovic, P. P. Orth, M. S. Scheurer, R. Prozorov, arXiv:2110.02025.

\bibitem{TRS1} G. M. Luke, Y. Fudamoto, K. M. Kojima, M. I. Larkin, J. Merrin, B. Nachumi, Y. J. Uemura, Y. Maeno, Z. Q. Mao, Y. Mori, H. Nakamura, and M. Sigrist, Nature 394, 558-561 (1998).

\bibitem{TRS2} G. M. Luke, A. Keren, L. P. Le, W. D. Wu, Y. J. Uemura, D. A. Bonn, L. Taillefer, and J. D. Garrett, Phys. Rev. Lett. 71, 1466 (1993).

\bibitem{TRS3} Y. Aoki, A. Tsuchiya, T. Kanayama, S. R. Saha, H. Sugawara, H. Sato, W. Higemoto, A. Koda, K. Ohishi, K. Nishiyama, and R. Kadono, Phys. Rev. Lett. 91, 067003 (2003).

\bibitem{HF1} G. M. Luke, Y. Fudamoto, K. M. Kojima, M. I. Larkin, B. Nachumi, Y. J. Uemura, J. E. Sonier, Y. Maeno, Z. Q. Mao, Y. Mori, D. F. Agterberg, Physica B 289-290, 373-376 (2000).

\bibitem{SheurerGenRelation} M. S. Scheurer, Phys. Rev. B \textbf{93} 174509 (2016).

\bibitem{topo0}  J. Alicea, Reports on Progress in Physics, \textbf{75}, 7 (2012).

\bibitem{DiodeThSgt} B. Zinkl, K. Hamamoto, M. Sigrist, arXiv:2111.05340.

\bibitem{DiodeTh} H. D. Scammell, J.I.A. Li, M. S. Scheurer, arXiv:2112.09115.

\bibitem{DiodeExp} J.-X. Lin, P. Siriviboon, H. D. Scammell, S. Liu, D. Rhodes, K. Watanabe, T. Taniguchi, J. Hone, M. S. Scheurer, J.I.A. Li, arXiv:2112.07841.


\bibitem{NbC_TaC} T. Shang , J. Z. Zhao, D. J. Gawryluk, M. Shi, M. Medarde, E. Pomjakushina, and T. Shiroka, Phys. Rev. B 101, 214518 (2020).

\bibitem{VC_CrC} R. Zhan and X. Luo, J. Appl. Phys. 125, 053903 (2019).

\bibitem{MoC} A. Huang, A. D. Smith, M. Schwinn, Q. Lu, T. R. Chang, W. Xie, H. T. Jeng, and G. Bian, Phys. Rev. Materials 2, 054205 (2018).



\bibitem{SCScS1} A. R. Moodenbaugh, D. C. Johnston, and R. Viswanathan, Mat. Res. Bull. \textbf{9}, 1671 (1974).

\bibitem{SCScS2} A. R. Moodenbaugh, D. C. Johnston, R. Viswanathan, R. N. Shelton, L. E. DeLong, and W. A. Fertig, Journal of Low Temperature Physics \textbf{33}, 175 (1978).

\bibitem{SI} See Supplementary Information for comprehensive calculation of various superconducting and normal state parameters.

\bibitem{McMillanMain} W. L. McMillan, Phys. Rev. 167, 331 (1968).

\bibitem{GapRatio} R. Combescot, G. Varelogiannis, Solid State Communications 93, 113 (1995).

\bibitem{Muon} A. D. Hillier, S. J. Blundell, I. McKenzie, I. Umegaki, L. Shu, J. A. Wright, T. Prokscha, F. Bert, K. Shimomura, A. Berlie, H. Alberto and I. Watanabe, Nat. Rev. Meth.Primers 2, 474 (2022)

\bibitem{TF1} M. Weber, A. Amato, F. N. Gygax, A. Schenck, H. Maletta, V. N. Duginov, V. G. Grebinnik, A. B. Lazarev, V. G. Olshevsky, V. Yu. Pomjakushin, S. N. Shilov, V. A.Zhukov, B. F. Kirillov, A. V. Pirogov, A. N. Ponomarev, V. G. Storchak, S. Kapusta, and J. Bock, Phys. Rev. B 48, 13022 (1993).
 
\bibitem{TF2} A. Maisuradze, R. Khasanov, A. Shengelaya, and H. Keller, J. Phys.: Condens. Matter 21, 075701 (2009).

\bibitem{TF_sig1} B. S. Chandrasekhar and D. Einzel, Ann. Phys., Lpz. 2, 535 (1993).

\bibitem{TF_sig2} R. Prozorov and R. W. Giannetta, Supercond. Sci. Tech-nol. 19, R41 (2006).

\bibitem{TF_sig3} D. T. Adroja, A. Bhattacharyya, M. Telling, Y. Feng, M. Smidman, B. Pan, J. Zhao, A. D. Hillier, F. L. Pratt, and A. M. Strydom, Phys. Rev. B 92, 134505 (2015).

\bibitem{2G_TF} D. A. Mayoh, A. D. Hillier, K. G\"otze, D. McK. Paul, G. Balakrishnan, and M. R. Lees, Phys. Rev. B 98, 014502 (2018).

\bibitem{ZF_KT} R. S. Hayano, Y. J. Uemura, J. Imazato, N. Nishida, T. Yamazaki, and R. Kubo, Phys. Rev. B 20, 850 (1979).

\bibitem{s+-} Y. Bang, and G. R. Stewart, J. Phys.: Condens. Matter 29, 123003 (2017).



\bibitem{FirstPrinciple} D. Shrivastava, S. P. Sanyal, Computational Condensed Matter \textbf{21}, e00418 (2019).

\bibitem{TunnelDiodeRes} C. T. Van Degrift, Rev. Sci. Instrum. \textbf{46}, 599 (1975).

\bibitem{LaNi} Arushi, D. Singh, A. D. Hillier, M. S. Scheurer, and R. P. Singh, Phys. Rev. B \textbf{103}, 174502 (2021).

\bibitem{Andrey} S. Maiti and A. V. Chubukov, Phys. Rev. B \textbf{87}, 144511 (2013).

\bibitem{BaKFeAs} V. Grinenko, R. Sarkar, K. Kihou, C. H. Lee, I. Morozov, S. Aswartham, B. B\"uchner, P. Chekhonin, W. Skrotzki, K. Nenkov, R. H\"uhne, K. Nielsch, S. -L. Drechsler, V. L. Vadimov, M. A. Silaev, P. A. Volkov, I. Eremin, H. Luetkens, and H.-H. Klauss, Nat. Physics \textbf{16}, 789 (2020).

\bibitem{spisFields} J. Garaud and E. Babaev, Phys. Rev. Lett. \textbf{112}, 017003 (2014).

\bibitem{BW} R. Balian and N. R. Werthamer, Phys. Rev. \textbf{131}, 1553 (1963).

\bibitem{GolubovMazin} A. A. Golubov and I. I. Mazin, Phys. Rev. B \textbf{55}, 15146 (1997).

\bibitem{DisorderSOCFu} K. Michaeli and L. Fu, Phys. Rev. Lett. \textbf{109}, 187003 (2012).

\bibitem{OurDisorderSOC} M. S. Scheurer, M. Hoyer, and J. Schmalian, Phys. Rev. B \textbf{92}, 014518 (2015).

\bibitem{BrydonScattering} D. C. Cavanagh, and P. M. R. Brydon, Phys. Rev. B \textbf{101}, 054509 (2020).

\bibitem{PdTeScattering} E. I. Timmons, S. Teknowijoyo, M. Konczykowski, O. Cavani, M. A. Tanatar, Sunil Ghimire, Kyuil Cho, Yongbin Lee, Liqin Ke, Na Hyun Jo, S. L. Bud'ko, P. C. Canfield, Peter P. Orth, Mathias S. Scheurer, and R. Prozorov, Phys. Rev. Research \textbf{2}, 023140 (2020).

\bibitem{Jonathan} D. Dentelski, V. Kozii, and J. Ruhman, Phys. Rev. Research \textbf{2}, 033302 (2020).

\bibitem{Ando2012} M. Kriener, Kouji Segawa, Satoshi Sasaki, and Yoichi Ando, Phys. Rev. B \textbf{86}, 180505(R) (2012).


\bibitem{Ando2014} Superconductor derived from a topological insulator heterostructure, Satoshi Sasaki, Kouji Segawa, and Yoichi Ando, Phys. Rev. B \textbf{90}, 220504(R) (2014).

\bibitem{Welp} M. P. Smylie, K. Willa, H. Claus, A. Snezhko, I. Martin, W.-K. Kwok, Y. Qiu, Y. S. Hor, E. Bokari, P. Niraula, A. Kayani, V. Mishra, and U. Welp, Phys. Rev. B \textbf{96}, 115145 (2017).


\bibitem{Brian} Clara N. Brei{$\clock$}, P. J. Hirschfeld, and Brian M. Andersen, Phys. Rev. B \textbf{105}, 014504 (2022).

\bibitem{DisorderCuprates} Li, ZX., Kivelson, S.A. and Lee, DH, npj Quantum Mater. \textbf{6}, 36 (2021).

\bibitem{BrydonRelation} P. M. R. Brydon, S. Das Sarma, H.-Y. Hui, and J. D. Sau, Phys. Rev. B \textbf{90}, 184512 (2014).

\bibitem{SrPtAs} P. K. Biswas, H. Luetkens, T. Neupert, T. St\"urzer, C. Baines, G. Pascua, A. P. Schnyder, M. H. Fischer, J. Goryo, M. R. Lees, H. Maeter, F. Br\"uckner, H.-H. Klauss, M. Nicklas, P. J. Baker, A. D. Hillier, M. Sigrist, A. Amato, and D. Johrendt, Phys. Rev. B \textbf{87}, 180503(R) (2013).


\bibitem{TaS2} David Dentelski, Ezra Day-Roberts, Turan Birol, Rafael M. Fernandes, and Jonathan Ruhman, Phys. Rev. B \textbf{103}, 224522 (2021).

\bibitem{KondoScreening} Ruan, W., Chen, Y., Tang, S. et al. Evidence for quantum spin liquid behaviour in single-layer 1T-TaSe2 from scanning tunnelling microscopy, Nat. Phys. \textbf{17}, 1154-1161 (2021). 


\end{thebibliography}

\begin{thebibliography}{References}

\bibitem{BG1} G. Grimvall, \textit{The Electron-Phonon Interaction in Metals} (North-Holland, Amsterdam, 1981).

\bibitem{BG2} A. Bid, A. Bora, and A. K. Raychaudhuri, Phys. Rev. B 74, 035426 (2006).

\bibitem{DL} D. A. Mayoh, J. A. T. Barker, R. P. Singh, G. Balakrishnan, D. McK. Paul, and M. R. Lees
Phys. Rev. B 96, 064521 (2017).

\bibitem{WHH1} E. Helfand, and N. R. Werthamer, Phys. Rev. 147, 288 (1966)
  
\bibitem{WHH2} N. R. Werthamer, E. Helfand, and P. C. Hohenberg, Phys. Rev. 147, 295 (1966).

\bibitem{Pauli1} A. B. Karki, Y. M. Xiong, I. Vekhter, D. Browne, P. W. Adams, D. P. Young, K. R. Thomas, Julia Y. Chan, H. Kim, and R. Prozorov, Phys. Rev. B 82, 064512 (2010).
 
\bibitem{Pauli2} J. K. Bao, J. Y. Liu, C. W. Ma, Z. H. Meng, Z. T. Tang, Y. L. Sun, H. F. Zhai, H. Jiang, H. Bai, C. M. Feng, Z. A. Xu, and G. H. Cao, Phys. Rev. X 5, 011013 (2015). 
 
\bibitem{Maki} K. Maki, Phys. Rev. B 148, 362 (1966).

\bibitem{Coh_Leng} M. Tinkham, Introduction to Superconductivity, 2nd ed. (McGraw-Hill, New York, 1996).

\bibitem{pene} T. Klimczuk, F. Ronning, V. Sidorov, R. J. Cava, and J. D. Thompson, Phys. Rev. Lett. 99, 257004 (2007).

\bibitem{McMillan} W. L. McMillan, Phys. Rev. 167, 331 (1968).

\bibitem{lowjump} C. B. Vining, R. N. Shelton, H. F. Braun, and M. Pelizzone, Phys. Rev. B 27, 2800 (1983).

\bibitem{multigap} B.Joshi, A. Thamizhavel, and S. Ramakrishnan, Phys. Rev. B 84, 064518 (2011).

\bibitem{SH1} O. J. Taylor, A. Carrington, and J. A. Schlueter, Phys. Rev. Lett. 99, 057001 (2007).

\bibitem{SH2} X. Xu, B. Chen, W. H. Jiao, Bin Chen, C. Q. Niu, Y. K. Li, J. H. Yang, A. F. Bangura, Q. L. Ye, C. Cao, J. H. Dai, Guanghan Cao, and N. E. Hussey, Phys. Rev. B 87, 224507 (2013).

\bibitem{SH3} C. Q. Niu, J. H. Yang, Y. K. Li, B. Chen, N. Zhou, J. Chen, L. L. Jiang, B. Chen, X. X. Yang, C. Cao, J. Dai, and X. Xu, Phys. Rev. B 88, 104507 (2013).


\bibitem{ApproxGap} A. Carrington, and F. Manzano, Physica C: Superconductivity, 385, 205 (2003).

\end{thebibliography}
\end{document}